\documentclass[useAMS,usenatbib]{mn2e}
\usepackage{graphicx}

\title{Timing observations of RRAT J1819-1458 at Urumqi Observatory}
\author[A. Esamdin et al]{A. Esamdin,$^{1}$\thanks{E-mail:
aliyi@uao.ac.cn} C. S. Zhao,$^{1,2}$ Y. Yan$^{1,2}$, N. Wang$^{1}$, H. Nizamidin$^{3}$, Z. Y. Liu$^{1}$\\
$^{1}$Urumqi Observatory, National Astronomical Observatories, CAS,
40-5 South Beijing Road, Urumqi, 830011, China\\
$^{2}$Graduate University of Chinese Academy of Sciences, Beijing
100049, China\\
$^{3}$Department of Physics, Xinjiang University , Urumqi, 830046,
China\\}
\begin{document}

\date{Received ?  / Accepted ?}

\pagerange{\pageref{firstpage}--\pageref{lastpage}} \pubyear{?}

\maketitle

\label{firstpage}

\begin{abstract}
{We report the timing-analysis results obtained for RRAT
J1819-1458 from regular timing observations at 1.54 GHz using the
Urumqi 25 m radio telescope between 2007 April to 2008 March. RRAT
J1819-1458 is a relatively young and highly magnetized neutron
star discovered by its sporadic short bursts in the Parkes
Multibeam Pulsar Survey data. In 94 hrs of observation data, we
detected a total of 162 dispersed bursts of RRAT J1819-1458 with
the signal-to-noise ratios (S/Ns) above 5-$\sigma$ threshold.
Among them, 5 bursts clearly show two-component structure. The S/N
of the strongest burst is 13.3. The source's DM measured through
our data is 196.0 $\pm$0.4 pc cm$^{-3}$. The timing position,
frequency and its first derivative were determined using standard
pulsar timing techniques on the arrival times of these individual
bursts. The accuracy of the solved rotating parameters are
improved comparing with that in previous publication. Our timing
position with 2-$\sigma$ error is consistent with the position of
its X-ray counterpart CXOU J181934.1-145804. The effect of timing
noise and the phase fluctuation of the individual short bursts on
the timing residuals is briefly discussed. The distribution of the
timing residuals is bimodal, which cannot be explained readily by
timing irregularity.}
\end{abstract}


\section{Introduction}

Recently, a remarkable new class of radio transient sources, the ``
Rotating Radio Transients''(RRATs) was discovered in a search for
isolated radio bursts in the Parkes Multibeam Pulsar Survey data
\citep{b15}. A total of 11 RRATs have been detected so far; they are
characterized by short radio bursts of a typical from 2 to 30 ms
duration, and the average time intervals between bursts range from 3
min to 3 hrs. Although the radio bursts of RRATs are sporadic, the
timing analysis of the bursts of 10 RRATs indicates that they are
likely to be rotating neutron stars with the periods ranging from
0.4 to 7 sec \citep{b15}.

RRAT J1819-1458 is the brightest and the most prolific
radio-burster of all 11 RRATs. At 1.4 GHz, the source is
characterized by dispersed radio bursts of average duration 3 ms
with one burst detected about every 3 min; its dispersion measure
(DM) is 196$\pm$3 pc cm$^{-3}$; and the peak flux of the brightest
burst detected so far is 3.6 Jy. The distance inferred from its DM
and position using the free electron density model of \citet{b2}
is about 3.6 kpc. Although the source is not detectable in
standard periodicity searches, the analysis of the spacings
between the bursts reveals a spin period, $P$ = 4.263 s, and a
spin period derivative, $\dot P$ = $576\times 10^{-15}$ s
s$^{-1}$. RRAT J1819-1458 has a characteristic age of 117 kyr, and
a relatively high inferred magnetic field strength of $5\times
10^{13} $ gauss \citep{b15}.

On the $P$-$\dot P$ diagram, the source is located in the same area
occupied by the high magnetic field radio pulsars, which suggests a
possible association between RRATs and the magnetars \citep{b15,
b24} or the X-ray dim isolated neutron stars \citep{b16,b5}. The
$Chandra$ observation detected the X-ray counterpart of RRAT
J1819-1458 \citep{b17}. Lately, using the $XMM-Newton$, \citet{b14}
discovered the X-ray pulsations with the period predicted by the
timing of radio bursts. The X-ray properties show that RRAT
J1819-1458 is a cooling neutron star, and unlikely to be a magnetor
\citep{b17,b16,b14,b19}.

Based on a study of subpulse modulation of pulsars \citep{b22},
\citet{b23} note that PSR B0656+14 has very energetic and sporadic
radio pulses with the same characteristic as RRATs, and suggest
that it could have been identified as a RRAT, were it not so
nearby. They point out that these energetic bursts are shown to be
unlike giant pulses \citep{b6,b10}, giant micropulses
\citep{b9,b8} or the pulses of normal pulsars \citep{b18,b11}.
\citet{b21} note RRATs are not directly related to nulling pulsars
which are characterized by turning on and off of the pulsations.
Although several models have been suggested to explain RRAT
phenomenon \citep{b25,b12,b4,b13}, the nature of emission of RRATs
is as yet unclear.

We observed RRAT J1819-1458 at 1540 MHz by using the Urumqi 25m
radio telescope from 10 April 2007 to 29 March 2008. In this paper,
we present the timing result of RRAT J1819-1458 based on these
observations. The details of the observations are presented in
Section 2. The process of burst detection is given in Section 3. The
timing process and results are described in Section 4. A brief
discussion is given in Section 5. In Section 6 we summarize this
study.

\section{Observations}
%%%%%%%%%%%%%%%%%%%%%%%%%%%%%%%%%%%%%%%%%%%%%%%%%%%%%%%%%%%%%%
\begin{figure*}
% \vspace{302pt}
 \includegraphics[width=110mm,height=150mm,angle=-90]{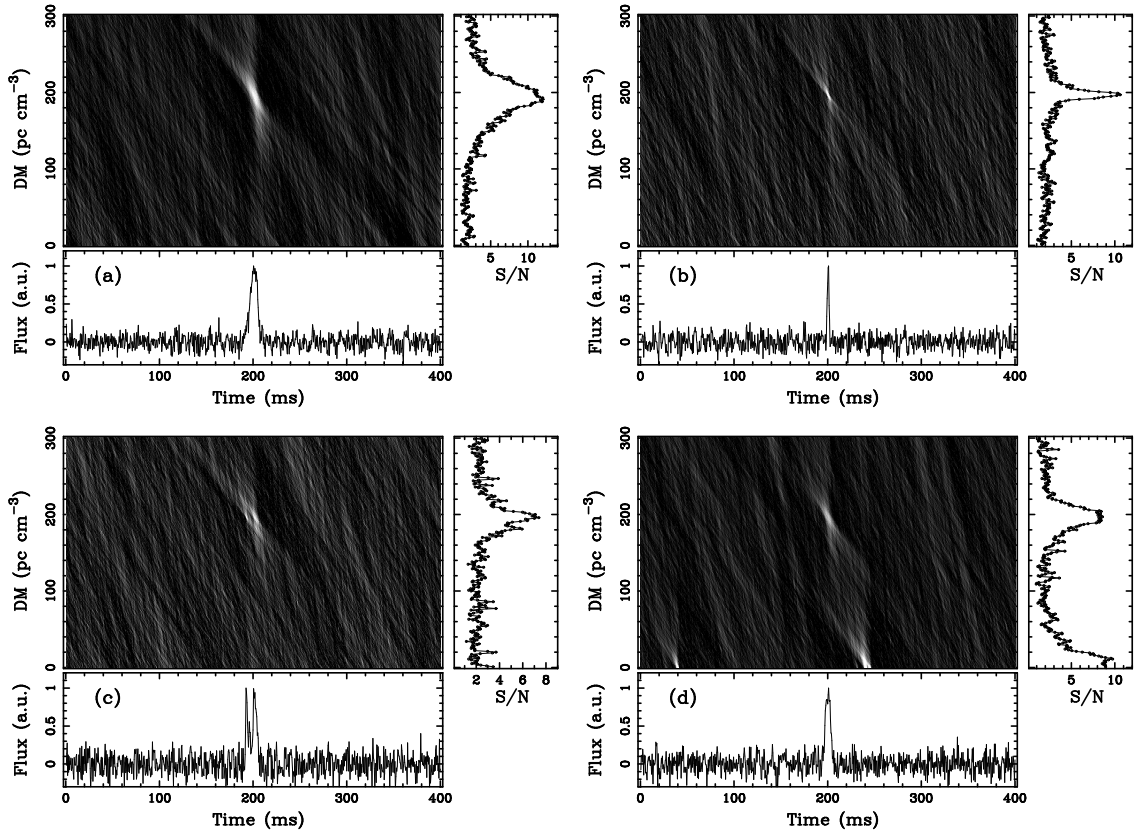}
 \caption{Diagnostic plots show four bursts detected in this work. Each diagnostic
 plot includes a DM versus time diagram in gray scale (top left panel), a S/N versus DM diagram (top right panel)
 and a dedispersed burst time series (bottom panel). Dedispersion procedure is performed from
zero DM to DM = 300 pc cm$^{-3}$ in steps of 1 pc cm$^{-3}$ to
obtain the DM versus time space. The S/N for each DM is calculated
for the maximum amplitude in a 400-ms dedispersed time series
centered on the candidate pulse. The bursts are clearly visible in
the DM versus time and the S/N versus DM diagrams with a maximum
S/N near the DM of 196 pc cm$^{-3}$. Plots (a), (b), and (c)
present a broad burst, a narrow burst and a bimodal burst
respectively. In addition to a real burst signal at the nominal
DM, plot (d) also shows two strong pulses of terrestrial radio
interference around the zero DM at about 40 ms and 243 ms,
respectively.}
\end{figure*}
%%%%%%%%%%%%%%%%%%%%%%%%%%%%%%%%%%%%%%%%%%%%%%%%%%%%%%%%%%%%%%

The observations of RRAT J1819-1458 have been regularly carried
out by using the Urumqi 25-m radio telescope with a dual-channel
cryogenic receiver that receives orthogonal linear polarizations
at the central observing frequency of 1540MHz. The receiver noise
temperature is less than 10 K. After mixing down to an
intermediate frequency, the two polarizations are each fed into a
filter bank of 128 contiguous channels, each of width 2.5 MHz. The
outputs from the channels are then square-law detected, filtered
and one-bit sampled at 0.5 ms interval \citep{b26}. The data
streams of all 256 channels are written to disk for subsequent
off-line processing. The start time of each observation is also
recorded to calculate the site arrival time of each pulse of
burst.

We detected the isolated dispersed bursts of RRAT J1819-1458 with
the S/N above 5-$\sigma$ threshold. The minimum peak flux density of
the detected signal is given by
\begin{equation}
 S_{min}= \frac{2\alpha\beta k(T_{rec}+T_{spl}+T_{sky})}{\eta A \sqrt{n_{p}\tau\Delta{f}}}
\label{eq:LebsequeI}
\end{equation}
where $\alpha = 5$ is the threshold signal-to-noise ratio, $\beta$
= $\sqrt {\pi/2}$ is a factor accounting for losses due to 1-bit
digitization, $k$ is Boltzmann's constant, $T_{rec}$, $T_{spl}$
and $T_{sky}$ are receiver, spillover and the sky noise
respectively ($T_{rec}+T_{spl}+T_{sky} \sim$ 32 K), $n_{p}$ = 2 is
number of polarizations summed, $\tau$=0.5 ms is the sampling
interval, $\Delta {f}$ = 320 MHz is the observing bandwidth, $\eta
\sim$ 57\% is the efficient of the antenna at 1540 MHz and
$\emph{A}$ = 490.87m$^{2}$ is the area of the antenna. We
calculated a minimum detectable pulse amplitude of $\sim$ 3.4 Jy
for our 5-$\sigma$ detection threshold.

In this work, a total of 47 observations were made within 25
observing sessions from 10 April 2007 to 29 March 2008. The time
span of each observation was 2 hrs. The sampling interval of all
observations was fixed to 0.5 ms. In total, 94 hrs of observing data
were collected in the one-year time span.

\section{burst detection}
\label{sect:data} Radio waves propagating through ionized plasma in
the interstellar medium experience a frequency-dependent delay due
to the dispersive effects of the plasma. In order to detect short
bursts from celestial sources, this frequency-dependent delay has to
be removed. The difference in arrival times, $\Delta t$ (ms),
between a pulse received at a high frequency, $f_{h}$, and a lower
frequency, $f_{l}$, is given by

\begin{equation}
 \Delta t= 4.148808\times DM\times
 (\frac{1}{f_{l}^{2}}-\frac{1}{f_{h}^{2}})
\label{eq:LebsequeI}
\end{equation}
where DM is the dispersion measure in pc cm$^{-3}$, and the
frequency values are in GHz. This equation was used to calculate the
signal delays in our 128 observing channels of each polarization.

The data processing was performed in several steps in order to
detect the dispersed short bursts from RRAT J1819-1458. The
observing data were first dedispersed by delaying successive
channels in time corresponding to the nominal dispersion measure DM
= 196 pc cm$^{-3}$, and then searched for all pulses above a
5-$\sigma$ S/N threshold in the dedispersed time series. Secondly,
we checked the frequency evolution of these candidate signals. The
dispersion is seen as a quadratic sweep across the observing
frequency band for the strong bursts detected in this study if two
polarizations summed. However, the weaker signals ( S/N $\leq$ 6.5
in our case ) are not strong enough to clearly present this
dispersed feature through all frequency channels. We therefore
checked if they were broadband by displaying the dedispersed time
series in 8 bands each of 16$\times$2.5MHz wide. Thirdly, the
dedispersion procedure was performed from zero DM to DM = 300 pc
cm$^{-3}$ in steps of 1 pc cm$^{-3}$ in order to confirm the
dispersion signature in a DM-time space. The S/N of the maximum
pulse amplitude was also computed for each DM over a time series of
400 ms centered on the candidate signal. From this analysis, a
diagnostic plot (as shown in Fig.1) was generated for each burst
candidate. These plots were then subjected to a careful visual
inspection to discriminate real bursts from spurious signals.

Fig. 1 shows the diagnostic plots of four bursts detected in this
work. Each diagnostic plot contains a DM versus time (top left
panel), a S/N versus DM (top right panel) and a dedispersed time
series diagram (bottom panel). In DM-time space, a burst is detected
as a vertical strip of pulses at multiple DMs around the nominal DM.
The strip is gradually broadening and finally smears out with the DM
either increasing or decreasing from the nominal DM. The bursts are
clearly visible in the DM versus time and the S/N versus DM diagrams
with maximum S/N around DM $\sim$ 196 pc cm$^{-3}$. Plots (a), (b),
and (c) in Fig. 1 show a broad, a narrow and a bimodal burst
respectively.

Some of our observing data were contaminated by Radio Frequency
Interference (RFI). The frequency-dispersion property of signals may
assist to distinguish between signals of celestial origin and
locally generated impulsive RFI \citep{b3}. As shown in plot (d) in
Fig. 1, the two high S/N pulses detected near the zero DM are likely
to be impulsive terrestrial interference. The two RFI signals spread
from zero DM (at about 40 ms and 243 ms respectively) to several
nearby DMs in the DM versus time diagram.

By using the searching method mentioned above, a total of 162
strong bursts, with the S/Ns ranging from 5 to 13.3, are detected
in 94 hrs of data. Among them, there are 5 bursts clearly show
two-component (bimodal) structure with the separations of the two
components ranging from 6 ms to 16.5 ms.

In order to determine the DM of this source through our data, we
produce the average S/N versus DM diagram (as shown in Fig. 2)
obtained by averaging all S/Ns in each DM step of all 162 bursts.
In this case, the dedispersion procedure for each burst is
performed from DM=96 to DM = 296 pc cm$^{-3}$ in steps of 0.1 pc
cm$^{-3}$. Fig. 2 shows that the peak of the distribution of the
average S/Ns is right at DM=196 pc cm$^{-3}$. The rms of S/N
fluctuation calculated from the average S/Ns in the DM range from
266 to 296 pc cm$^{-3}$ is 0.032. To determine the DM error,
3-$\sigma$ level of the S/N fluctuation ($\sim 0.1$) is taken. The
DM error measured according to the S/N distribution is $\pm$0.2 pc
cm$^{-3}$. We take twice the value as the error of the DM, and
obtain the improved DM = 196.0 $\pm$0.4 pc cm$^{-3}$ (listed in
Table 1).

%%%%%%%%%%%%%%%%%%%%%%%%%%%%%%%%%%%%%%%%%%%%%%%%%%%%%%%%%%%%%%
\begin{figure}
% \vspace{302pt}
 \includegraphics[width=51mm,height=70mm,angle=-90]{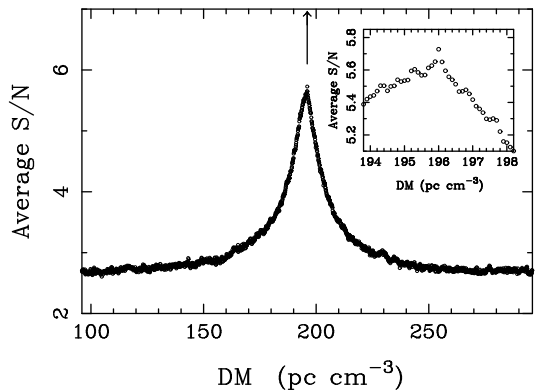}
 \caption{Average S/N versus DM diagram obtained by averaging all S/Ns
 in each DM step of all 162 bursts. The dedispersion procedure for each burst is performed
from DM=96 to DM = 296 pc cm$^{-3}$ in steps of 0.1 pc cm$^{-3}$.
The peak of the S/N distribution is right at DM = 196 pc cm$^{-3}$
as indicated by the vertical arrow. The inserted plot shows the
variation of the average S/Ns in DM range from 193.8 to 198.2 pc
cm$^{-3}$.}
\end{figure}
%%%%%%%%%%%%%%%%%%%%%%%%%%%%%%%%%%%%%%%%%%%%%%%%%%%%%%%%%%%%%%%

\section{Timing analysis and results}
\citet{b15} note that RRAT J1819-1458 is only detectable through
individual radio bursts. The timing process of the source is
therefore slightly different from the standard pulsar timing process
in which the times of arrival (TOAs) are determined by fitting a
template profile to the observed mean pulse profiles. We measured
position, frequency, $\nu$, and frequency first derivatives, $\dot
\nu$, using standard pulsar timing techniques on the arrival times
of individual bursts of the source.

The data are dedispersed (at DM = 196 pc cm$^{-3}$) relative to
the central observing frequency to form the pulses of the 162
bursts. In our case, TOA of a pulse of burst refers to the pulse's
midpoint which is determined by the center of a Gauss fit to the
pulse. However, as mentioned in above section, 5 of the 162 bursts
present two component structure. In case of these
bimodal-structure bursts, the midpoints of stronger components are
used to determine their TOAs. TOAs of the pulses at the telescope
are then measured by projecting the start times of the
observations to these midpoints. The uncertainties of TOAs are
calculated by dividing the half power full widths ($W_{50}$) of
the pulses by their signal to noise ratios. TOAs are processed
using the standard TEMPO 2 software package \citep{b7}, which
first converts them to solar system barycentric TOAs at infinite
frequency using the Jet Propulsion Laboratory Solar-system
ephemeris DE405 \citep{b20}, and then performs a least-square fit
to determine the model parameters. The position, $\nu$, and $\dot
\nu$ are quoted for an epoch (MJD 54400) near the midpoint of our
data span. The timing residuals, which are the differences between
the actual pulse arrival times and those calculated from the
fitted model, are then obtained to investigate the phases of
bursts and the source's rotational behavior.

%%%%%%%%%%%%%%%%%%%%%%%%%%%%%%%%%%%%%%%%%%%%%%%%%%%%%%%%%%%%%%
\begin{figure}
% \vspace{302pt}
 \includegraphics[width=55mm,height=70mm,angle=-90]{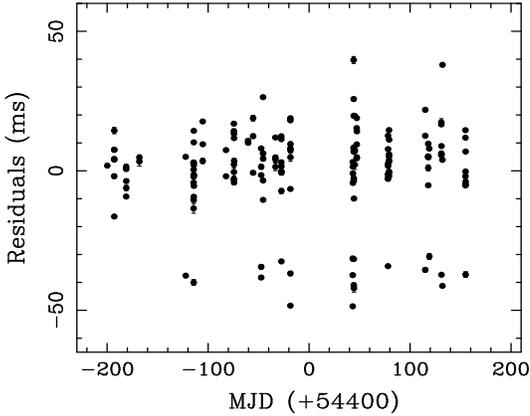}
 \caption{Timing residuals for RRAT J1819-1458 after fitting for the position,
 rotational frequency and its first derivative. The residuals of 162 individual
 bursts range from -48.57 ms to 39.71 ms. The rms residual is 16.14 ms.
 It is about 0.4\% of the source's period.}
\end{figure}
%%%%%%%%%%%%%%%%%%%%%%%%%%%%%%%%%%%%%%%%%%%%%%%%%%%%%%%%%%%%%%%

Table 1 lists the best-fitting parameters and their uncertainties,
including position, $\nu$ and $\dot \nu$ at epoch MJD 54400.
Uncertainties in the last digit quoted are given in parentheses.
These uncertainties are taken to be twice the standard errors
obtained from TEMPO2. Table 1 also lists the DM determined through
our data, the epoch of the period, the number of TOAs included in
the timing solution, the MJD range covered and the rms of the post
fit timing residuals. We also calculated characteristic age,
$\tau_c=P/(2\dot P)$, the surface dipole magnetic field strength,
$B_s=3.2\times10^{19}(P\dot P)$, the rate of loss in rotational
energy, $\dot E=4\pi^2I\dot P/P^3$ (using a standard neutron star
moment of inertia $I=10^{45} gcm^2$), and the magnetic field
strength at the light cylinder, $B_{Lc}=9.35\times 10^5\dot
P^{1/2}P^{-5/2}$, as listed in Table 1. These derived parameters
are very close to those presented by \citet{b15}. The timing
residuals of the 162 bursts are shown in Fig. 3. In this work, the
MJD range of the available TOAs is two times longer than for
\citet{b15}, and hence the accuracy of the timing solutions has
improved correspondingly.

%%%%%%%%%%%%%%%%%%%%%%%%%%%%%%%%%%%%%%%%%%%%%%%%%%%%%%%%%%%%%%
\begin{table}
    \caption{Timing parameters and derived parameters for RRAT
    J1819-1458. Uncertainties in the last digit quoted are given in
parentheses. The uncertainties of timing parameters are twice the
standard TEMPO errors.}
    \begin{tabular}{lcc}
    \hline
    \hline
        \ Paramters                     &Values\\
        \hline
         Right Ascension (J2000)        & 18:19:33.8(4)\\
        Delination.(J2000)             &-14:58:01(16)\\
        $\nu $ ($s^{-1}$)                   &0.2345648893(7)\\
        $\dot \nu $ ($10^{-14}s^{-2}$)           &-3.104(2)\\
                                               \\
       DM (pc cm$^{-3}$)                   &196.0(4)\\
                                                  \\
        Epoch (MJD)                    &54400\\
        Number of TOAs                 &162\\
        Time span (MJD)                &54200.1 -- 54555.1\\
        Rms residual (ms)                     &16.14\\
                                                    \\
        $\tau_c$  (kyr)                        &119.81(8)\\
      $B_{s}$ (gauss)                           &4.963(2)$\times $10$^{13}$\\
       $\dot E$  (erg s$^{-1}$)            &2.876(2)$\times $10$^{32}$\\
      $B_{Lc}$ (gauss)                           &5.818(2)\\
    \hline
    \end{tabular}
\end{table}
%%%%%%%%%%%%%%%%%%%%%%%%%%%%%%%%%%%%%%%%%%%%%%%%%%%%%%

Fig. 4 compares the position obtained in this work with that of
published positions. Our fitted position at the reference epoch
with J2000 coordinates is: right ascension =
18$^{h}$19$^{m}$33.$^{s}$8$\pm $ 0$^{s}$.4 and declination =
-14$^{h}$58$^{h}$01$^{h} \pm $16 (J2000). This is consistent with
the more accurate position for the X-ray counterpart (CXOU
J181934.1-145804) published in \citet{b17} and with the timing
position reported in \citet{b15} assuming 2 $\sigma $
uncertainties for the values quoted in that paper. Long-term
observations are necessary to acquire more precise timing position
of the source.

As shown in Fig.3, the measured timing residuals range from -
48.57 ms to 39.71 ms. The rms residual (16.14 ms) is about 0.4\%
of the source's period. RRAT J1819-1458 is a relatively young
neutron star which is very likely to show a significant timing
irregularity. The residuals of single-pulse timing are contributed
both by the timing irregularity and by the modulation of pulse
phase in the radiation window. It is therefor difficult to
discriminate the intrinsic rotating irregularity (especially the
timing noise) from the phase modulation through the timing
residuals of the sporadic individual bursts. However, Fig.3 shows
that no significant glitch, which would produce continually
growing (negative) arrival-time residuals, is detected in the one
year of rotating history of the source.

%%%%%%%%%%%%%%%%%%%%%%%%%%%%%%%%%%%%%%%%%%%%%%%%%%%%%%%%%%%%%%
\begin{figure}
 %\vspace{302pt}
   \includegraphics[width=70mm,height=70mm,angle=-90]{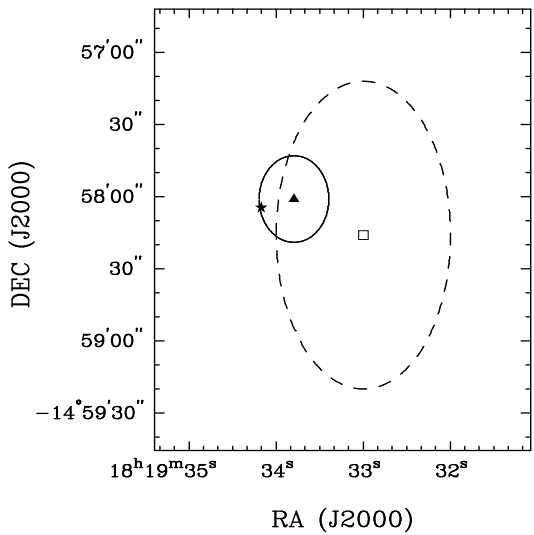}
 \caption{A comparison between the timing position obtained in this work (filled triangle)
 with that reported in \citet{b15} (open square) and with the
 position of CXOU J181934.1-145804 \citep{b17} (star).
 The error ellipses indicate the 2-$\sigma $ errors of the standard tempo uncertainties.}
\end{figure}
%%%%%%%%%%%%%%%%%%%%%%%%%%%%%%%%%%%%%%%%%%%%%%%%%%%%%%%%%%%%%%

Fig. 5 presents a histogram of residuals and a S/N versus residual
diagram. As shown in Fig. 3 and Fig. 5, the residual distribution of
these strong bursts is clearly bimodal, peaked at about -37 ms
(leading peak) and 3 ms (main peak) respectively.  The pulse number
detected around the leading peak and the main peak are 18 and 144
respectively. The implication of the distribution of the timing
residuals is discussed in following section.

%%%%%%%%%%%%%%%%%%%%%%%%%%%%%%%%%%%%%%%%%%%%%%%%%%%%%%%%%%%%%
\begin{figure}
% \vspace{302pt}
   \includegraphics[width=70mm,height=70mm,angle=-90]{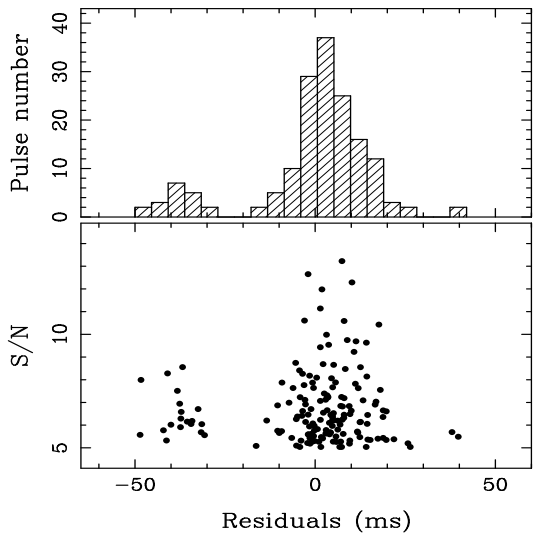}
 \caption{Histogram of residuals (top panel), and the S/N versus residuals diagram (bottom panel).
 The distribution of the timing residuals of 162 bursts shows a bimodal structure, peaked at about -37 ms and 3 ms.
 This may indicate a possible weak preceding component 40 ms ahead of the main
 component in the radiation window of RRAT J1819-1458. The S/Ns of the bursts range from 5 to 13.3.}
\end{figure}

 %%%%%%%%%%%%%%%%%%%%%%%%%%%%%%%%%%%%%%%%%%%%%%%%%%%%%%%%%%%%%%

\section{Discussion}

In order to estimate a possible contribution of stochastic timing
noise to the timing residuals, we calculated the cumulative phase
contributed by timing noise over time interval of our observations
($\sim$ 355 days). If we express the rotation frequency as a
Taylor series, the pulse phase at time $t$ is given by
\begin{equation}
 \phi= \phi_0+\nu t+\frac{1}{2} \dot \nu t^{2}+\frac{1}{6}\ddot \nu
 t^3  \cdots
\label{eq:LebsequeI}
\end{equation}
where $\phi_0$ is the phase in cycles at time $t=0$. The fourth term
in this equation can be used to estimate the cumulative phase
contributed by stochastic timing noise over time interval $t$. We
therefore carried out a timing solution with the second frequency
derivative $\ddot \nu$ included, and detected no significant
decrease of the rms residual. Because $|\ddot \nu|<2\sigma_{\ddot
\nu}$ in our case, following \citet{b1}, we quoted an upper limit of
the cumulative phase ($\delta_t$) as
\begin{equation}
\delta_t<\frac{1}{6\nu} (2\sigma_{\ddot \nu})t^3
\label{eq:LebsequeI}
\end{equation}
where $\nu$ = 0.23456 is the rotating frequency, $\sigma_{\ddot
\nu}$ = 1.19$\times $10$^{-24}$ is the formal uncertainty of the
second frequency derivative, and $t\sim$ 30672000 s is the time
interval over our observations. We obtained the cumulative phase,
$\delta_t<$ 50 ms, in a one year time span.

Although the cumulative phase reaches 50 ms in a one year
duration, the result of a fitting for $\nu$ and $\dot \nu$ to a
simple rotating model with the $\ddot \nu$ included shows that the
rms residual is less than 2 ms. The rms of timing residuals in
Table 1 is much larger than that contributed by the stochastic
timing noise expected for this source over one year. If we take a
typical duty circle value of 3\% of normal pulsars, the
burst-phase span of a 4.26 sec pulsar is expected to be about 130
ms. For these reasons, we may note that the residuals in Fig. 3
are strongly dominated by random fluctuations in the burst phases,
and can be good representations of the phases of the individual
bursts in the source's radiation window.

The bimodal distribution of the residuals (as shown in Fig. 5)
cannot be explained readily by timing irregularity, and may
indicate the possible existence of a more sporadic burst activity
in the longitudinal region preceding the main radiation region.
However, the number of detected bursts is not yet sufficient to
make this a firm conclusion. More observations and an accurate
phase alignment are needed to obtain a reliable luminosity
distribution of the radiation window.

As shown in Table 1, RRAT J1819-1458 is a relatively young pulsar
with a long spin period and a high $B_s$. In fact, there are only
16 pulsars in ANTF Pulsar Catalog with the measured $B_s$ values
higher than that of this source. It has a modest $\dot E$ and a
low $B_{Lc}$ in the known pulsar population. The unusual radio
emission and the detection of X-ray pulsation of RRAT J1819-1458
make it a very interesting object. Its sporadic radio activity may
has similar mechanism with one of the two known radio-emission
phenomena, i.e. giant-pulse emission in some pulsars \citep{b6,
b10} or the very sporadic strong pulses detected in PSR B0656+14
\citep{b23}. The known giant-pulse emitters have very high values
of $B_{Lc}$ ($> 10^{5}$ gauss) and $\dot E$ ($\sim 10^{36 - 38}$)
which are suggested to be indicators of giant-pulse emissivity
\citep{b27, b30, b10}. Therefor, the low $B_{Lc}$ and $\dot E$ of
RRAT J1819-1458 is against the suggestion of giant-pulse origin of
its radio bursts \citep{b15, b14}. The characteristic ages of PSR
B0656+14 and RRAT J1819-1458 are similar, and the X-ray pulsation
characteristics of RRAT J1819-1458 seems to be similar with that
of PSR B0656+14 \citep{b14}. However, we should note that the two
sources have very different spin period, $B_s$ and $\dot E$. The
anomalous X-ray pulsars (AXPs) have very high values of $B_s$.
Therefor, the transient radio emission detected in AXPs XTE
J1810-197 and 1E 1547.0-5408 \citep{b28, b29} may also has a
similar mechanism with the radio emission of this high-$B_s$ RRAT.
However, the radio emission characteristics of the two AXPs are
quite different from that of RRAT J1819-1458 \citep{b15}. By
studying the X-ray properties, \citet{b14} note that RRAT
J1819-1458 could be a transition object between the pulsar and
magnetar. More studies are needed to determine the reason of the
source's unusual radio behavior. A study of individual-burst
properties of RRAT J1819-1458 mainly based on the bursts presented
in this paper will be reported in a subsequent paper.

\section{Summary}
 \label{sect:discussion}
We have carried out timing observations of RRAT J1819-1458 by
using the Urumqi 25 m radio telescope at a center frequency of
1.54 GHz. A total of 47 observtions were made in 25 sessions from
10 April 2007 to 29 March 2008. In total 162 strong bursts, with
S/N $\geq$ 5, were detected through a careful burst-detection
processing. 5 bimodal bursts, with the component separations
ranging from 6 ms to 16.5 ms, were noted. We obtained a more
precise DM = 196.0 $\pm$0.4 pc cm$^{-3}$ from our data. We have
presented the improved timing position, $\nu$ and $\dot \nu$ for
RRAT J1819-1458 based on these bursts using standard pulsar timing
techniques on the arrival times of individual bursts. The location
of the X-ray counterpart (CXOU J181934.1-145804) is now within
2-$\sigma$ error ellipse of our timing position.

No evidence of significant glitch and timing noise were found in
the one year of rotating history of this source, likely due to the
phase modulation of the sparse individual pulses of bursts within
its radiation windows. Continued monitoring is needed to
investigate its rotating irregularity. We found a bimodal
distribution of the timing residuals, which may suggest a possible
two-component structure of the source's radiation window, with a
more sporadic component leading the main one. However, more
observations are also necessary to confirm this conclusion.

\section*{Acknowledgments}

We thank the reviewer for the very helpful comments. The authors
are supported by the National Natural Science Foundation of China
(NSFC) under grant 10573026 and 10778631, the program of the Light
in China's Western Region(LCWR) under grant. LHXZ200602, and the
Knowledge Innovation Program of the Chinese Academy of Science
under grant KJCX2-YW-T09.

\label{lastpage}


\begin{thebibliography}{99}

\bibitem[\protect\citeauthoryear{Arzoumanian et al.}{1994}]{b1}
Arzoumanian Z., Nice D. J., Taylor J. H., Thorsett S. E., 1994 , ApJ, 422, 671

\bibitem[\protect\citeauthoryear{Camilo et al.}{2007}]{b29}
Camilo F., Ransom S. M., Halpern J. P., Reynolds J., 2007, ApJ,
666, L93

\bibitem[\protect\citeauthoryear{Camilo et al.}{2006}]{b28}
Camilo F., Ransom S. M., Halpern J. P., Reynolds J., Helfand D.
J., 2006, Nature, 442, 892

\bibitem[\protect\citeauthoryear{Cognard et al.}{1996}]{b27}
Cognard I., Shrauner J. A., Taylor J. H., Thorsett S. E., 1996,
ApJ, 457, L81

\bibitem[\protect\citeauthoryear{Cordes \& Lazio}{2004}]{b2}
Cordes J. M., Lazio T. J. W., 2004, ASPC, 317, 211

\bibitem[\protect\citeauthoryear{Cordes \& Mclaughlin}{2003}]{b3}
Cordes J. M., Mclaughlin M., 2003, ApJ, 596, 1142

\bibitem[\protect\citeauthoryear{Cordes \& Shannon}{2006}]{b4}
Cordes J. M., Shannon R. M., 2006, astro-ph/0605145

\bibitem[\protect\citeauthoryear{for
details about XDINSs see Haberl}{2004}]{b5} Haberl F, 2004, AdvSpR,
33, 638

\bibitem[\protect\citeauthoryear{e.g. Hankins et al.}{2003}]{b6}
Hankins T. H., Kern J. S., Weatherall J. C., Eilek J. A., 2003,
Nature, 422, 141

\bibitem[\protect\citeauthoryear{Hobbs, Edwards \& Manchester}{2006}]{b7}
Hobbs G. B., Edwards R. T., Manchester R. N., 2006, MNRAS, 369, 655

\bibitem[\protect\citeauthoryear{Johnston \& Romani}{2002}]{b8}
Johnston S., Romani R., 2002, MNRAS, 332, 109

\bibitem[\protect\citeauthoryear{e.g. Johnston et al.}{2001}]{b9}
Johnston S., van Straten W., Kramer M., Bailes M., 2001, ApJ, 549,
L101

\bibitem[\protect\citeauthoryear{Knight et al.}{2006}]{b10}
Knight H. S., Bailes M., Manchester R. N., Ord S. M., Jacoby B. A.,
2006, ApJ, 640, 941

\bibitem[\protect\citeauthoryear{Kramer et al.}{2003}]{b11}
Kramer M., Karastergiou A., Gupta Y., Johnston S., Bhat N. D. R.,
Lyne A. G., 2003, A\&A, 407, 655

\bibitem[\protect\citeauthoryear{Li}{2006}]{b12}
Li X. D., 2006, ApJ, 646, L139

\bibitem[\protect\citeauthoryear{Luo \& Melrose}{2007}]{b13}
Luo Q., Melrose D., 2007, MNRAS, 378, 1481

\bibitem[\protect\citeauthoryear{Mclaughlin \& Cordes}{2003}]{b30}
McLaughlin M. A., Cordes J. M., 2003, ApJ, 596, 982

\bibitem[\protect\citeauthoryear{Mclaughlin et al.}{2006}]{b15}
McLaughlin M. A., Lyne A. G., Lorimer D. R., Kramer M., Faulkner
A. J., Manchester R. N., Cordes J. M., Camilo F., Possenti A.,
Stairs I. H., Hobbs G., D'Amico N., Burgay M., O'Brien J. T.,
2006, Nature, 439, 817

\bibitem[\protect\citeauthoryear{Mclaughlin et al.}{2007}]{b14}
McLaughlin M. A., Rea N., Gaensler B. M., Chatterjee S., Camilo F.,
Kramer M., Lorimer D. R., Lyne A. G., Israel G. L., Possenti A.,
2007, ApJ, 670, 1307

\bibitem[\protect\citeauthoryear{Popov, Turolla \& Possenti}{2006}]{b16}
Popov S. B., Turolla R., Possenti A., 2006, MNRAS, 369, 23

\bibitem[\protect\citeauthoryear{Reynolds et al.}{2006}]{b17}
Reynolds S. P., Borkowski K. J., Gaensler B. M., Rea N.; McLaughlin
M., Possenti A, Israel G., Burgay M., Camilo F., Chatterjee S.,
Kramer M., Lyne A. G., Stairs I., 2006, APJ, 639, 71

\bibitem[\protect\citeauthoryear{e.g. Ritchings}{1976}]{b18}
Ritchings R. T., 1976, MNRAS, 176, 249

\bibitem[\protect\citeauthoryear{Rutledge}{2006}]{b19}
Rutledge R. E., 2006, astro-ph/0609200

\bibitem[\protect\citeauthoryear{Standish}{2004}]{b20}
Standish, E. M. 2004, A\&A, 417, 1165.

\bibitem[\protect\citeauthoryear{Wang, Manchester \& Johnston}{2007}]{b21}
Wang N., Manchester R. N., Johnston S., 2007, MNRAS, 377, 1383

\bibitem[\protect\citeauthoryear{for details about this system see Wang et al.}{2001}]{b26}
Wang N., Manchester R. N., Zhang J., Wu X. J., Yusup A., Lyne A. G.,
Cheng K. S., Chen M. Z., 2001, MNRAS, 328, 855

\bibitem[\protect\citeauthoryear{Weltevrede, Edwards \& Stappers}{2006}]{b22}
Weltevrede P., Edwards R. T., Stappers B. W., 2006, A\&A, 445, 243

\bibitem[\protect\citeauthoryear{Weltevrede et al.}{2006}]{b23}
Weltevrede P., Stappers B. W., Rankin J. M., Wright G. A. E., 2006,
ApJ, 645, 149

\bibitem[\protect\citeauthoryear{for details about magnetar see Woods \& Thompson}{2006}]{b24}
Woods P. M., Thompson C., 2006, in Lewin W. H. G., van der Klis M.,
eds, Compact stellar X-ray sources.  Cambridge Univ. Press,
Cambridge, 547

\bibitem[\protect\citeauthoryear{Zhang, Gil \& Dyks}{2007}]{b25}
Zhang B., Gil J., Dyks J., 2007, MNRAS, 374, 1103

\end{thebibliography}
\end{document}